\documentclass[]{aa}
\usepackage{supertabular}
\usepackage{longtable}
\usepackage{txfonts}
\usepackage{graphicx}
\def\Swift{\emph{Swift}}

\def\til{\ensuremath{\sim\,}}
\def\sqiglt{\hbox{\rlap{\lower.55ex \hbox {$\sim$}}\kern-.05em \raise.4ex \hbox{$<$}\,}}
\def\sqiggt{\hbox{\rlap{\lower.55ex \hbox {$\sim$}}\kern-.05em \raise.4ex \hbox{$>$}\,}}

\begin{document}

\title{Accurate early positions for {\it Swift\/} GRBs: enhancing X-ray
 positions with UVOT astrometry.}

\author{M.R. Goad\inst{1}, L.G. Tyler\inst{1}, A.P. Beardmore\inst{1},
 P.A. Evans\inst{1}, S.R. Rosen\inst{1}, J.P. Osborne\inst{1},
 R.L.C. Starling\inst{1}, F.E. Marshall\inst{2}, V. Yershov\inst{3},
 D.N. Burrows\inst{4}, N. Gehrels\inst{2}, P. Roming\inst{4},
 A. Moretti\inst{5}, M. Capalbi\inst{6}, J.E. Hill\inst{7} J. Kennea\inst{4},
 S. Koch\inst{4}, D. Vanden Berk\inst{4},}

\institute{Department of Physics and Astronomy, University of Leicester,
LE1 7RH, UK.
\and NASA Goddard Space Flight Center, Greenbelt, MD 20771, USA.
\and Mullard Space Science Laboratory, Department of Space and Climate Physics, University College London, Holmbury St. Mary, Dorking, Surrey, RH5 6NT, UK,
\and Pennsylvania State University, University Park, PA 16802, USA
\and INAF-Osservatorio Astronomica di Brera, via Bianchi 46, 23807
\and ASI Science Data Center, Via Galileo Galilei, I-00044 Frascati,
Italy.
\and Universities Space Research Association, 10211 Wincopin Circle, Suite
500, Columbia, MD 21044-3432, USA}

\date{Received : / Accepted : }

\abstract{The {\it Swift\/} Gamma Ray Burst satellite routinely provides
prompt positions for GRBs and their afterglows on timescales of a few hundred
seconds. However, with a pointing accuracy of only a few arcminutes, and a
systematic uncertainty on the star-tracker solutions to the World Coordinate
System of 3--4 arcseconds, the precision of the early XRT positions is limited
to 3--4 arcseconds at best.  This is significant because operationally, the
XRT detects $>$95\% of all GRBs, while the UVOT detects only the optically
brightest bursts, $\sim$30\% of all bursts detected by BAT; thus early and
accurate XRT positions are important because for the majority of bursts they
provide the best available information for the initial ground-based follow-up
campaigns.

Here we describe an autonomous way of producing more accurate prompt XRT
positions for GRBs and their afterglows, based on UVOT astrometry and a
detailed mapping between the XRT and UVOT detectors. The latter significantly
reduces the dominant systematic error -- the star-tracker solution to the
World Coordinate System.  This technique, which is limited to times when there
is significant overlap between UVOT and XRT PC-mode data, provides a factor of
2 improvement in the localisation of XRT refined positions on timescales of
less than a few hours. Furthermore, the accuracy achieved is superior to
astrometrically corrected XRT PC mode images at early times (for up to 24
hours), for the majority of bursts, and is comparable
to the accuracy achieved by astrometrically corrected X-ray positions based on
deep XRT PC-mode imaging at later times.

\keywords{gamma-ray: bursts -- Gamma-rays, X-rays: afterglows }}

\titlerunning{UVOT-enhanced XRT positions for {\it Swift\/} detected GRBs.}
\authorrunning{Goad et al.}
\maketitle

\section{Introduction}

Since its launch in November 2004, the {\it Swift\/} Gamma-Ray Burst Explorer
(Gehrels et~al. 2004) has been routinely observing the prompt gamma-ray and
early afterglow emission of Gamma-Ray Bursts (GRBs) in the astrophysically
important minutes to hours timescale after the onset of the burst. One of the
main science goals of {\it Swift\/} is the fast dissemination of GRB positions
for dedicated ground and space-based follow-up. In this context, accuracy is
of paramount importance, particularly if prompt spectroscopic observations of
bright, optical afterglows is desired.

As part of its routine service to the GRB science community, the Burst Alert
Telescope (hereafter BAT, Barthelmy et al. 2004), and the X-ray Telescope
(hereafter XRT, Burrows et~al. 2004) routinely provide, via the Gamma-Ray
Burst Coordinates Network (GCN), automated on-board positions on timescales of
less than a few minutes, and their respective science teams provide refined
positions (after ground-based re-processing) on timescales of minutes--hours.

For the majority of GRBs the XRT is the only narrow-field instrument on board
{\it Swift\/} to provide a localised source position, that is, the majority of
{\it Swift\/}-detected bursts are optically-faint, being either intrinsically
faint, reddened, or at high redshift (see e.g. Roming et al. 2006), and are
therefore below the detector sensitivity of the UltraViolet and Optical
Telescope (hereafter UVOT, Roming et al. 2005) .

The XRT prompt positions have a precision of order 3.5 arcseconds (systematic)
with an additional statistical uncertainty based on the number of observed
counts in the initial image (statistical error, $\epsilon = 22.63\times
counts^{-0.48}$, see Hill et~al. 2004). In the high-count regime, the majority
of the uncertainty derives from the uncertainty in the attitude reconstruction
provided by the star-trackers and uncertainties in the alignment of the
boresight to the star-tracker (see e.g. Moretti et~al. 2006). Conversely, for
the UVOT, knowledge of the boresight and star-tracker solutions are of lesser
importance, as, for any image with long enough exposure, UVOT images may be
astrometrically corrected by matching positions of detected sources relative
to those in stellar catalogues (e.g. USNO-B1) and applying an astrometric
correction to the image. For a single UVOT image, the typical residual mean
error in source positions, after applying astrometric corrections, is of order
0.5--1.0 arcseconds, depending on the number and location of sources in the
17'x17' field of view of UVOT. Thus it is only for those rare occasions when
there are insufficent stellar matches to provide an accurate astrometric
solution, that knowledge of the UVOT boresight is essential.
% We confine a detailed discussion of the UVOT
% boresight and verification of the star-tracker performance to an appendix.

By contrast our inability to provide accurate astrometric corrections to early
XRT observations means that the uncertainty in the XRT positions places severe
limitations on the ability of large aperture ground-based telescopes to
perform early (within a few hours) spectroscopic observations of the GRB
afterglow and its environment.

To improve this situation, we have undertaken an investigation to ascertain
whether the UVOT can be used as a Super Star-Tracker, ie. to provide the
necessary aspect information to astrometrically correct XRT prompt positions,
thereby providing GRB localisations with an accuracy and precision of better
than a few arcseconds. This is particularly important if the GRBs are
optically faint, and no obvious counterpart is seen in UVOT.  Our
investigation has a number of aims : (i) estimate the stability of the optical
bench on which the XRT and UVOT are mounted, by determining within a single
orbit, the accuracy with which XRT and UVOT track individual sources in
detector coordinates, then if the tracking between XRT and UVOT proves
sufficiently stable, (ii) determine an accurate mapping between XRT and UVOT
detector coordinate systems, as a means of providing ultimately, (iii) XRT
refined positions enhanced by UVOT astrometric solutions.

For step (i) we work in detector coordinates, and verify that drifts in source
positions in detector coordinates within an orbit, caused by drift in the
spacecraft attitude, are accurately tracked in both instruments. If this
condition is satisfied, then in principal, XRT detector coordinates can be
mapped into UVOT detector coordinates (step ii), transformed into the world
coordinate system using the star-tracker solution, and then (step iii) aspect
corrected with respect to a standard stellar catalogue.

Such a process is feasible, if and only if the following conditions are met :

\begin{itemize}

\item{The pointing direction of the XRT and UVOT telescopes track each other
with high precision (ie. the flexure in the optical bench is minimal).}

\item{Time-dependent focusing variations in both XRT and UVOT, due to changes
in the mirror optics (for example, those caused by temperature variations) are
minimal.}

\item{There is sufficient temporal overlap between XRT PC mode and UVOT image
mode data.}

\end{itemize}

The paper is set out as follows: In \S2, we investigate the accuracy with
which XRT and UVOT track individual sources in detector coordinates. In \S\S3
\& 4 we outline a method for mapping XRT--UVOT detector coordinates. In \S5 we
describe in detail our implementation of an automated procedure for producing
UVOT-enhanced XRT positions in real-time. In \S6 we provide validation of our
fitting procedure, and our list of UVOT-enhanced XRT positions for 154
GRBs. Finally, the results are summarised in \S7.

\section{Tracking the alignment between UVOT and XRT.}

In order to track the UVOT/XRT alignment during the course of an orbit, we
have selected a target which has been observed in XRT PC event
mode\footnote{While the XRT has a number of different observing modes, only
Imaging and PC event mode data provide 2-dimensional images of the sky
necessary for determining GRB positions} and UVOT event mode, allowing us to
time-slice the observations into short enough intervals to track any drift in
the source position in detector coordinates, but with sufficient counts in the
XRT to allow accurate centroiding on the source. We require that the XRT PC
event mode data are not piled-up, ie. sources with count rates of less than
0.8 ct/s. The bright, radio-loud AGN, 3C 279 ($m_{\rm B}=17.8$), with an XRT
(0.2-10 keV) count-rate of $\sim0.3$~ct/s is a suitable target for this
analysis.

\begin{figure}
\resizebox{\hsize}{!}{\includegraphics[angle=0]{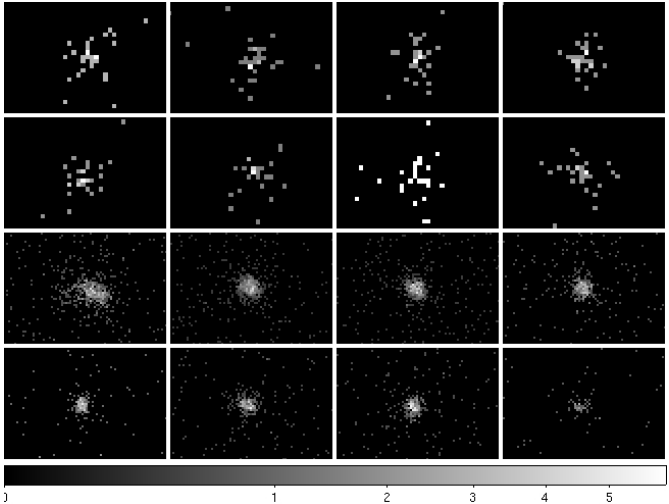}}
\caption{Top -- time-sliced (100~s) XRT PC mode detector coordinate
images (82''x54'') for the first orbit of XRT PC mode data. Bottom --
corresponding UVOT V-band detector coordinate images (4 in UVW2, followed by 4
in UVM2, each 40''x25'') spanning the same time intervals.  Note the
significant blurring of the source in the first panel of the UVOT data.}
\label{image_slices_xrt}
\end{figure}

\begin{figure}
\resizebox{\hsize}{!}{\includegraphics[angle=0]{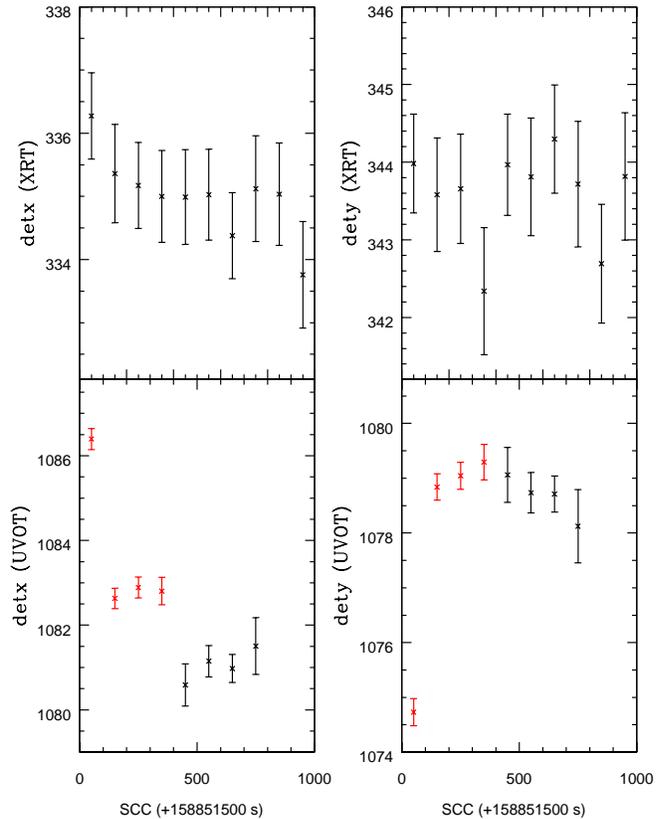}}
\caption{A comparison of the detector coordinate positions of the XRT (upper
 panels) and the UVOT (lower panels) as a function of Space Craft Clock (SSC)
 time for one orbit of observations of 3C279. The colours denote the different
 UVOT filters used (UVW2 -- red, UVM2 -- black). One XRT pixel is 2.36
 arcseconds to a side. One UVOT pixel is 0.5 arcseconds to a side. }
\label{detposorb}
\end{figure}

3C~279 has been observed on several occasions by {\it Swift\/} XRT as part of
its routine calibration program. This analysis uses data taken between
2006-01-13T00:37:00 and 2006-01-13T23:37:57. With 14 orbits of usable data
(15~ksec XRT PC mode), the X-ray data was time-sliced into 100~s bins,
providing 10 time slices per orbit, with approximately 30 ct/bin. For each
time slice we extracted an image in detector coordinates and measured the
source position by fitting the model PSF (the background is negligible over a
100s exposure). A sequence of 8 images each of 100~s duration is shown in
Figure~\ref{image_slices_xrt}, panels 1--8. Using the same time intervals we
then accumulate images from the UVOT event list
(Figure~\ref{image_slices_xrt}, panels 9-16). Because UVOT event data are
taken through several different filters (in this instance $\approx$ 3.8 ksec
in UVM2 and $\approx$ 5.9 ksec in UVW2), this typically allowed 7-8 images to
be constructed per orbit with overlapping XRT PC mode event data, half in UVW2
and half in UVM2. Figure~\ref{image_slices_xrt} shows that in the first UVOT
detector coordinate image, the source is significantly blurred.  This
blurring, which is also present in the first image of subsequent orbits, is
due to small residual movements in the spacecraft pointing direction
immediately following the slew. As a consequence large offsets in the UVOT
(and also XRT) detector coordinate positions can be recorded between images
taken during and after settling (see Figure~\ref{detposorb}). We note that in
some of the 3C279 UVOT event mode data there is evidence for significant
($>$1.0'') changes in the detector coordinate position of an individual source
due to filter changes within an orbit, though the occurrence of these is not
easily predicted.

\begin{figure}
\resizebox{\hsize}{!}{\includegraphics[angle=0]{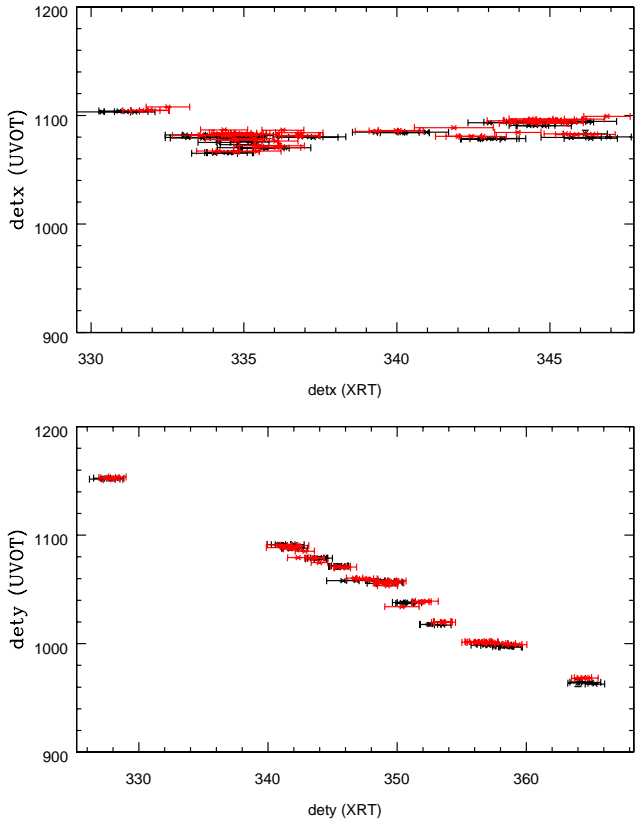}}
\caption{Correlation between the detector coordinate positions of 3C~279 
for XRT-PC and UVOT (UVW2 --red and UVM2 -- black) event mode data. One XRT pixel is 2.36
 arcseconds to a side. One UVOT pixel is 0.5 arcseconds to a side.}
\label{correlate}
\end{figure}

In Figure~\ref{correlate} we show the detector coordinate positions for the
XRT and the UVOT, for each of the 14 orbits of data (except orbit 13 which is
unusable) (approximately 100 positions). The corresponding positions are
highly correlated, particularly in the y-direction, for this particular roll
angle (121.6~degrees). The y-directions for the XRT and UVOT detector
coordinates are oriented 180+28.7 degrees relative to one another and are
therefore expected to show a strong negative correlation in their detector
dety coordinates, as is observed in Fig~\ref{correlate}.

%In figure 5 we show the source in detector
%coordinates for all 15 orbits of data for (a) XRT and (b) UVOT. Highlighted
%are the measured detx, dety positions for orbits 1, 7 and 15. 

The measured separation of the source position in detector coordinates between
orbits 1 and 15 is 192.6 UVOT pixels and 39.4 XRT pixels respectively. For a
pixel scale of 0.5 arcsec/pixel (UVOT) and 2.36 arcsec/pixel (XRT), this
separation corresponds to 96.3 arcsec (UVOT) and 92.9 arcsec (XRT), i.e. a
$<$4\% error in the tracking (consistent with the accuracy to which the XRT
pixel sizes are known). The difference in the angular direction
is consistent with the optical bench design which has UVOT
oriented 28.7 degrees relative to XRT along the y-axis, with their y-axes
running in opposite directions.

The accuracy with which source positions are tracked in both detectors
suggests that there is no evidence for flexure in the optical bench and that
the azimuth for each of the two detector coordinate planes is aligned to very
high precision.

%\begin{figure}
%\resizebox{\hsize}{!}{\includegraphics[angle=0]{./FIGS/xrt_pos_seq_im.ps}}
%\resizebox{\hsize}{!}{\includegraphics[angle=0]{./FIGS/ds9_uvot_detxyim.ps}}
%\caption{Time sequence showing the variation (in detector coordinates) of
%      the source position for XRT (upper panel) and UVOT (lower panel).}
%\label{}
%\end{figure}

\section{Mapping XRT to UVOT detector coordinates}

We start from the simplifying assumption that the detector plate scales for
both instruments are linearised and both lie in the x-y plane\footnote{Any
tilt in the azimuthal direction (ie. the angle between the normals to the two
detector planes) can manifest itself as a change in the ratio of the
plate-scales in both x and y directions. However we find no evidence for
this.}.  We fit the mapping between XRT and UVOT detector coordinates through
a combination of rotation, zero-point offset, and multiplicative plate-scale
(ratio of the size of UVOT to XRT detector pixels) by minimising the
$\Delta$detx, $\Delta$dety residuals between measured and predicted UVOT
detector positions.  That is, we minimise the difference between the predicted
UVOT detector position $U_{detx_{p}}$ and the measured UVOT detector position
$U_{m}$ (determined by fitting the model PSF), where $U_{detx_{p}}$,
$U_{dety_{p}}$ are given by :

\begin{equation}
\label{matrixeqn}
\left(
\begin{array}{l}
U_{detx_{p}} \\
U_{dety_{p}} 
\end{array}
\right)
=
\alpha
\left(
\begin{array}{c}
X_{detx} \\
X_{dety}
\end{array}
\right)
\left(
\begin{array}{cc}
\cos\theta & -\sin\theta \\
\sin\theta & \cos\theta
\end{array}
\right)
+
\left(
\begin{array}{c}
\delta \\
\epsilon
\end{array}
\right)
\end{equation}

\noindent where $\alpha$ is the ratio of the UVOT/XRT detector plate-scales,
$\theta$ is the rotation angle between the two detectors, 
and ($\delta$, $\epsilon$) are the zero point offsets. Note that the form of
this rotation matrix accounts for the $\approx 210$ degree difference in
orientation of the detector coordinate y-axes between XRT and UVOT.

Given the spacecraft pointing uncertainty (few arcminutes), the detector
coordinate map relating XRT detector coordinate to UVOT coordinate positions
must span a substantial fraction of the 17'x17' field of view of UVOT.  Thus
we require a large number of observations covering a large fraction of the
detector and spanning a large range in roll-angles (to reduce the likelihood
of producing a map which is accurate along one direction only, since repeated
pointings at similar roll angles tend to cover a very narrow area on the
detector plane, and to verify that roll-angle effects are absent). The 3C279
UVOT event mode data used previously to determine the tracking stability is
unsuitable on its own due to the restricted range in roll-angles covered by
these observations. Fortunately, there are numerous ($>$100) XRT PC-mode and
UVOT V-band observations of 3C~279 taken over a large range in roll-angle and
available within the public {\it Swift\/} data archive\footnote{The current
(as of April 2006) UVOT filter wheel sequence for a new GRB is: 10~s V-band
settling exposure, a 100~s White filter finding chart, 400~s V-band filter
finding chart, followed by observations through all UV and optical filters in
turn followed by a White filter and then V-band filter finding chart.  We
decided in the first instance to calibrate the map for the V-band filter
only. White has not been used in the calibration as yet because although these
observations begin once the as-settled flag is set, the spacecraft can still
show large drifts in attitude (lasting up to a few hundred seconds after the
GRB trigger), causing substantial drifts in the detector coordinate
position. Since detector to sky coordinate conversions are determined for a
specific instant in time, and GRBs generally fade rapidly (ie. most of the
X-ray photons arrive during the first few tens of seconds of the exposure) GRB
X-ray detector coordinate positions can often be inaccurate if the spacecraft
has not yet come to rest. For the majority of bursts V-band observations
start when drifts in the spacecraft attitude are far smaller, so the V-band
filter is the earliest to give an easily interpretable position.}.

Based on minimising the residuals between the predicted and measured UVOT
positions in detector coordinates using observations of 3C~279 taken over a
broad range in spacecraft roll angle and covering the central 8 arcminutes of
the XRT detector, the best-fitting model parameters are as follows : ($\theta
= 28.7072, \alpha = 4.69, \delta = 481.72, \epsilon= 3023.97$)
\noindent with a 90\% confidence limit of 1.3''. Both the orientation,
$\theta$, and ratio of the plate-scale, $\alpha$, are entirely consistent with
the known relative detector geometries for XRT and UVOT.

\section{PSF fitting of XRT PC-mode GRB prompt and early afterglow emission using the Cash-Statistic}

Aside from the systematic error derived from the XRT-UVOT detector mapping,
the precision with which a source position can be determined also depends upon
the precision with which the XRT detector coordinate position can be
determined, particularly in the low photon count regime when the statistical
error on the XRT position will dominate the error budget.

In order to produce as accurate a position as possible, we have employed PSF
fitting to determine GRB positions. The procedure is as follows: a list of
potential sources in the XRT detector coordinate image is found using a
specially written version of the {\it celldetect\/} algorithm (Harnden
et~al. 1984), which is tuned for the XRT PSF (Moretti et al. 2005), which at
the same time is able to discount any potential residual hot-pixels as
sources. For the exposures obtained during the UVOT/XRT data overlap
intervals (see below), which are typically 100-400~s, the background level is
negligible, and experience with the algorithm suggests a source can reliably
be detected with as few as 10 counts with a S/N of 2.0.  Once a source has
been detected its position is determined using a PSF centroid fitting
algorithm based on the maximum likelihood method of Cash (1979).  The standard
$C$-statistic from this paper is modified so that the integral of the PSF
probability distribution is renormalised in the presence of bad CCD detector
columns which have no exposure. Several such columns in the centre of the CCD
resulted from a micrometeoroid strike on 2005 May 27 (Abbey et~al. 2005).

\begin{figure*}
\resizebox{\hsize}{!}{\includegraphics[angle=0,width=16.0cm]{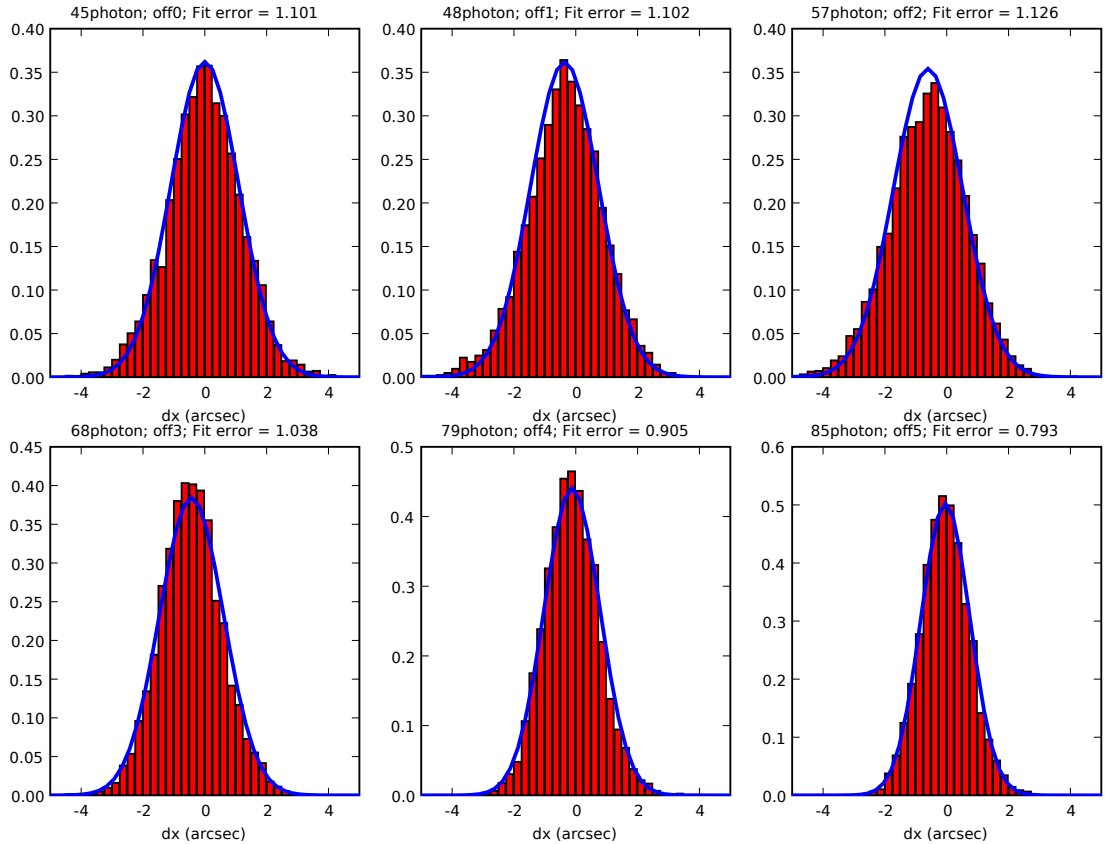}}

\caption{Distribution of residuals obtained in simulations of XRT images when
a non piled-up source is located near 5 bad columns.  From top-left to
bottom-right, the bad columns are systematically offset by, 0, 1, 2, 3, 4, 5
XRT pixels from the centre of the PSF. The red histograms show the
distribution of differences of the simulated positions from the input
position, while the blue curve is the Gaussian average fit error. The poorest
case (a 2 pixel offset) yields a 0.6 arcsecond mismatch which is added as an
additional systematic error when this offset occurs. There were 5000
simulations per run, for a source of 100 counts intensity.}

\label{monte}
\end{figure*}

The centroiding algorithm has been tested using Monte-Carlo simulations (see
Fig~\ref{monte}).  Simulating sources with different numbers of counts, photons
were drawn at random from the XRT PSF probability distribution and the
centroid determined. Comparison of the distribution of position residuals
obtained after 5000 simulations per run confirmed that the returned 90\%
statistical error, $\epsilon$, derived from the fit for non piled-up data was
accurate and followed the relation $\epsilon = 11.6~{counts}^{-0.5}$
arcseconds.

\begin{figure*}
\resizebox{\hsize}{!}{\includegraphics[angle=0,width=8.0cm]{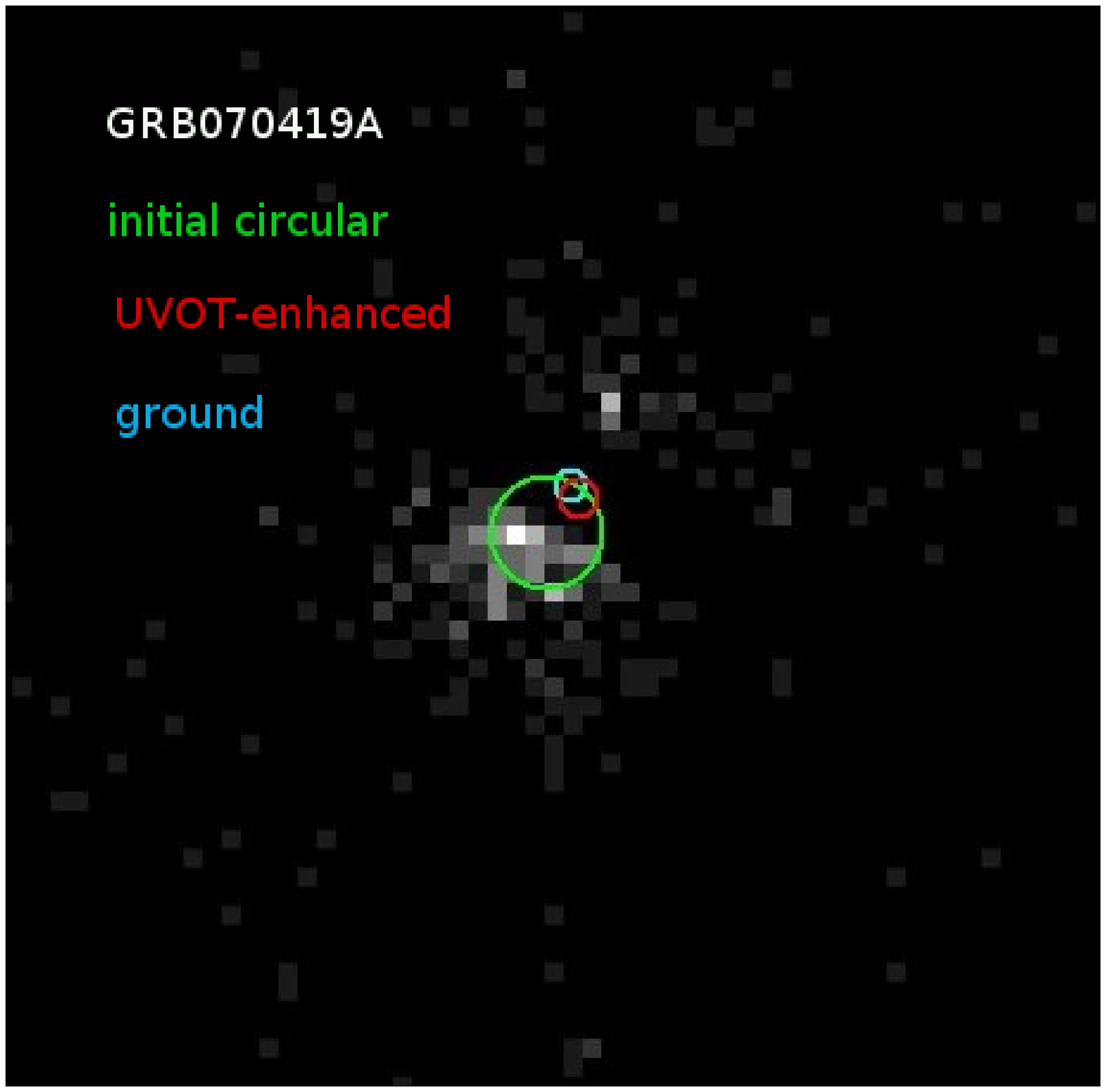}}

\caption{A 2.2 x 2.2 arcminute sky coordinate image of the XRT PC-mode
(1st orbit only) observations of GRB070419A, a GRB which by chance is centred
on the bad columns.  Also shown are the XRT refined position (large circle)
and the UVOT enhanced and ground-based optical positions, the small dark, and
small light circles respectively). PSF fitting accounting for bad columns,
provides improved position estimates and ultimately better final positions for
XRT detected GRBs.}

\label{grb070419a}
\end{figure*}

We have also verified that the centroiding performs well in the presence of
bad columns. At the time of writing, standard processing of XRT PC mode data
removes 5 columns centred at detector x-coordinate 292 and 3 columns at
320. Simulations were performed as before, but either 3 or 5 columns were
removed from the image at various offsets from the expected PSF centre. The
results showed that the centroids were unaffected by the presence of 3 bad
columns, while for 5 bad columns, a worse case residual offset of 0.6
arcseconds was seen when the true source centre was placed inside the bad
columns (Fig~\ref{monte}).  For comparison, a simple barycentre centroid
estimator applied to the 5 bad column simulations showed a worse case offset
of 2.4 arcseconds.  Based on these results, if a source centroid is found to
be inside the 5 bad columns an additional systematic error of 0.6 arcseconds
is included. In Figure~\ref{grb070419a} we show the XRT PC mode sky coordinate
image from the first orbit of data for GRB~070419A, a burst which was severely
affected by the presence of vetoed bad columns. Our UVOT-enhanced XRT
position, using the PSF fitting routine, accounting for bad columns and the
UVOT-XRT detector map, is a significant improvement on the XRT refined
position (Stratta 2007).

During the XRT's automated response to a GRB, the observed X-ray light curve
may be piled-up in Photon Counting mode when the observed count rate is
greater than $\sim 0.8 \thinspace {\rm count} \thinspace {\rm
s}^{-1}$. Because of this, we also tested how accurate the centroiding
algorithm is when a source is piled-up.  A number of observations were
selected from the {\it Swift} data archive which had different levels of
pile-up. For each observation a piled-up PSF profile was determined from the
image. The piled-up profile was used as the input PSF probability distribution
for simulating sources, which then had their centroids fitted using the non
piled-up PSF. The distribution of residuals showed that the position
statistical error, $\epsilon$, is underestimated when the number of counts in
an image is low and the data are piled-up. The simulations allowed a 90\%
confidence error correction term to be derived, to be added in quadrature to
the formal fit error, equal to $1.96\thinspace
{rate}^{1.2} \times {counts}^{-0.9}$, where $rate$ is the observed count rate
when greater than the non piled-up limit of $0.8 \thinspace {\rm count}
\thinspace {\rm s}^{-1}$.

% description of algorithm
% description of simulations and figures
% description of errors - and what to do with piled-up sources
% description of problems - bad pixels - bad columns 
% comment on failure due to drift in attitude (settling images) [Mike]
% comparison of errors with NATs formula - ie accuracy in low count limit

\section{Production of UVOT-enhanced XRT positions}
\label{sec:produce}

% trigger mechanism (of pipeline by new burst)
% time of availability (do we have any data for this yet?)
% list of products
% description of web-pages - table summary - couple of figures
% things to look out for
% why things can and do go wrong (ie catch-alls)
% web-address

The publicly available on-line quick-look data area is checked every 10
minutes for data associated with a new trigger number. Once a dataset
containing both PC mode XRT data and V-band UVOT imaging data is made
available by the {\it Swift\/} Data Center (SDC), the creation of a
UVOT-enhanced XRT position is initiated. The process is fully automated.  The
timescale for which an enhanced position is produced depends ultimately on the
timescale for which SDC processed down-linked data are delivered to the Data
Centre Quick-Look area. In most cases this will occur within a few hours of
the burst trigger, unless the spacecraft is an orbit which does not pass over
the Malindi ground station, during which time no data down-link is possible.

%Fig.~\ref{fig:postime}
%shows how soon after the trigger positions were produced for xx bursts. 90\% of%the time, a position was available within yy hr of the \Swift\ trigger.

The automated position enhancement procedure works on the first observation of
a new GRB, which is first broken down into times when PC mode X-ray data were
obtained simultaneously with UVOT V-band images. Currently our automated
procedure requires simultaneity of data since some drift in attitude is
expected, particularly during the spacecraft settling phase. It is possible
that this constraint may be relaxed in the future. In most cases\footnote{We
have derived UVOT-enhanced XRT positions for 84 out of 135 GRBs observed by
{\it Swift\/} since April 2006 when the current filter sequence was
implemented, and for 154 bursts in total (see Table~3).} there will be at
least one V-band image that overlaps with XRT-PC event mode data during the
first orbit of observations\footnote{Exceptions include : very bright bursts
for which XRT remains in WT mode for the whole of the first orbit
(e.g. GRB~051117A, Goad et~al. 2007), delayed slews (which have a variety of
causes, including Sun-Earth-Moon constraints), and bright stars in the UVOT
field of view.}.  Such times are referred to as \emph{overlaps\/} throughout
the rest of this paper. For each interval of overlap the following steps are
taken:

\begin{itemize}

\item{An XRT PC-mode detector co-ordinate image is produced from the cleaned
event lists, and any sources therein are automatically detected. The position
centroid of each detected source is determined using the PSF fitting routine
described in \S4.  If no source is found within 40" of the XRT TDRSS position
(if available), or in cases where there is no XRT TDRSS position, within the 3
arcminute BAT error circle, this overlap is rejected.

If an XRT TDRSS position is available, the source located closest to this
position is assumed to be the GRB. Otherwise, for the first three overlaps the
brightest source within the BAT position error circle is assumed to be the
GRB\footnote{Experience shows that for prompt (within a few minutes) slews,
the brightest source within the XRT field of view is the GRB afterglow. This
is not necessarily true for delayed slews, or for XRT ToO observations of GRBs
detected by other instruments.}; after this the weighted mean of these three
positions is calculated and the nearest source to this position is assumed to
be the GRB.}

\item{If the GRB position was determined using fewer than 10 XRT photons it is
deemed unreliable and the overlap is rejected.}

\item{If the source position is more than 40" away from the XRT TDRSS position
(when available) or from the weighted mean position (formed from the weighted
average of UVOT enhanced XRT positions determined within each of the first
three overlaps) of the UVOT-enhanced XRT position otherwise, it is flagged as
a non-GRB and the overlap is rejected.}

\item{The source position for the GRB in XRT detector coordinates is converted
to an equivalent detector coordinate position for UVOT, using the XRT-UVOT
detector coordinate mapping procedure described in \S4. The UVOT
detector position is then transformed into a sky coordinate position using the
\Swift\ tool {\sc swiftxform} which applies the aspect information supplied by
the star-trackers to the UVOT reference pixel.}

\item{Serendipitous source positions in the UVOT sky coordinate image are
matched with their corresponding optical counterparts in the USNO-B1
catalogue, using the \Swift\ tool {\sc uvotskycorr} which determines the
astrometric transformation required to correct UVOT sky-coordinate positions
relative to USNO-B1. Overlaps containing UVOT astrometric corrections of $>5$"
resulting from drifts in the spacecraft attitude during the settling phase
(which may last up to a few hundred seconds following a repointing of the
telescope) are rejected as unreliable.}

\item{Finally, the derived quaternions describing the transformation from the
UVOT WCS to USNO-B1 WCS are applied to the
XRT position, yielding for each overlap interval, an UVOT
astrometrically-enhanced XRT position for the GRB.}
\end{itemize}

The precision of the final position is formed from a combination of both
statistical and systematic errors. These include the statistical uncertainty
associated with the PSF fitting procedure (step 1), and the residual
uncertainty in the astrometric solution relating UVOT WCS to USNO-B1 WCS
(steps 5--6). These uncertainties are added in quadrature. From these we
determine a weighted mean position and error.  Any of the individual overlap
positions which lie outside of the weighted means' 3-$\sigma$ error circle are
rejected as outliers and the weighted mean position and error re-calculated.
Finally, the systematic uncertainty associated with the mapping between
XRT--UVOT detector coordinates (1.3", 90\% confidence, \S3) plus an additional
systematic uncertainty, necessary to place 90\% of optically detected GRBs
within our 90\% error circle (see \S6), are added in quadrature at the end.

\subsection{Dissemination of results}
\label{webpage}

Once determined, the UVOT-enhanced XRT position is made available to the GRB
community via an automated GCN circular. More detailed information is
published electronically at http://www.swift.ac.uk/xrt\_positions. This online
material contains a list of every enhanced position produced, links to
detailed results pages for individual bursts, and links to images of the UVOT
filter sequence; as well as descriptions of the process of deriving these and
other types of XRT position (e.g. SPER positions). The filter sequence images
are a graphical representation of the UVOT observations through each filter
and are provided both for completeness and diagnostic purposes.

\begin{figure*}
\resizebox{\hsize}{!}{\includegraphics[angle=-90,width=16.0cm]{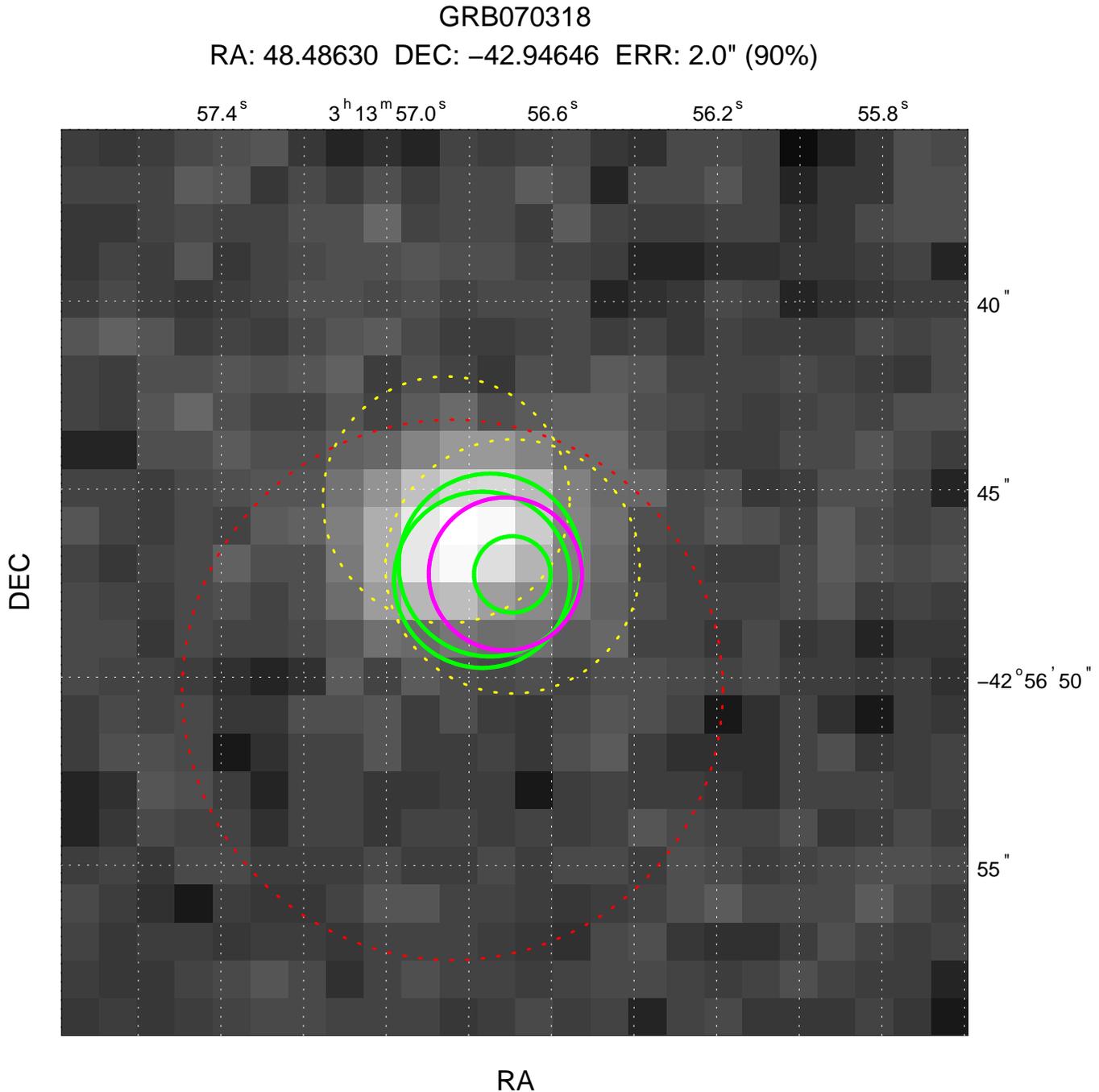}}

\caption{An example 25x25 arcsecond UVOT sky coordinate V-band image from the
detailed results page for GRB~070318 (see
http://www.swift.ac.uk/xrt\_positions/00271019/image.php for
details). The final position (magenta) is based on 3 overlaps (green). One
further overlap was rejected as the fit contains fewer than 10 photons (dashed
red), and two were rejected as the astrometric correction was considered too
large (dashed yellow).}

\label{astrom_ximage_all_narrow}
\end{figure*}

The detailed results pages can be accessed by appending
$<$targetID\footnote{the target ID is the trigger number padded to 8 digits
with leading zeroes. e.g. GRB 070531 was trigger 280958, so has target ID
00280958.}$>$/image.php to the above URL, and a page exists for each burst
with at least one overlap. This page gives the source position and error, the
number of overlaps included in the position determination and the total amount
of exposure time within those overlaps. We also provide a breakdown of the
number of overlaps excluded from the final position calculation and the reason
for their rejection (i.e low counts, large astrometric correction etc.).
These results are tabulated at the top of each page.

Below this, we provide two UVOT V-band images centred on the GRB position.
The first image is a close up image from the summed, astrometrically
corrected, UVOT V-band images overlaid with the enhanced XRT 90\% confidence
error circle. A larger version of this image is also provided, overlaid in
this case with the position and error circle from each overlap. Those
positions used to calculate the enhanced position are shown in green,
positions rejected due to low counts are indicated in red, while all other
rejected positions are in shown in yellow. Fig.~\ref{astrom_ximage_all_narrow}
shows one example, GRB~070318.

\begin{figure}
\resizebox{\hsize}{!}{\includegraphics[angle=-90,width=8.0cm]{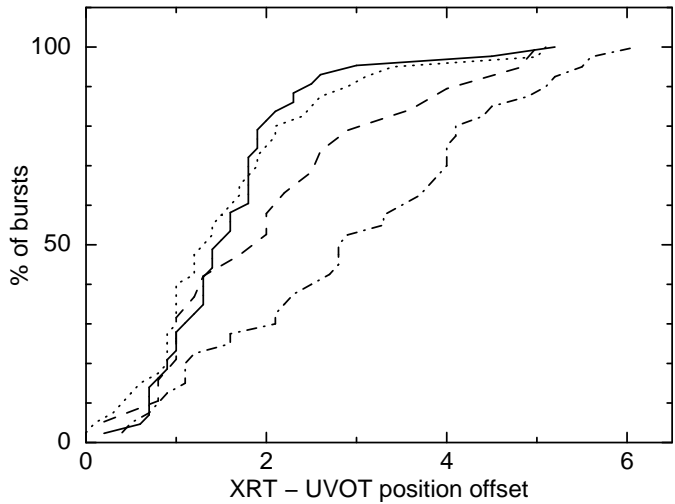}}
\caption{The cumulative distribution of XRT position offsets (in arcseconds)
with respect to 40 UVOT GRB positions available in the literature. Solid-line:
UVOT-enhanced XRT position offsets, the result of this work. Dot-dashed line:
'Refined' XRT position offsets, as derived in ground analysis and published in
GCN circulars and the Swift data table at
http://swift.gsfc.nasa.gov/docs/swift/archive/grb\_table/. Dashed line:
astrometrically corrected XRT positions at $\approx$24 hours (Butler 2007).
Dotted line : final astrometrically corrected positions from deep XRT images
(Butler 2007).}
\label{precise}
\end{figure}

\subsection{When a position cannot be determined}

An enhanced position cannot be determined for every burst. The most common
reason for failure is the absence of overlapping PC-mode XRT data and V-band
UVOT data. This can occur for example if the UVOT is in blocked mode, due to
the presence of a bright star within the UVOT field of view.  Alternatively,
if the slew was delayed or the burst initially faint, the number of source
counts at the location of the burst in XRT PC mode data may be less than the
10 counts required to reliably centroid on a source.  Cases for which no
position correction was possible are reported on the main results page.  If at
least one overlap exists, but no corrected position was produced, a detailed
results page is also produced, and a brief explanation of the reason for the
failure summarised on the main results page.

% sys.ps and nosys.ps

\subsection{Manual Intervention}

The automated procedure is not a catch-all in that incorrect positions may in
principle be made.  For example, if the initial XRT GCN position notice is
wrong, this could lead to an enhanced position for a non-GRB being
produced. The same is possible if there is no XRT position notice and the
brightest source in the BAT field of view at early times is not the GRB. Both
of these eventualities are very unlikely, but not impossible. Such cases will
be manually corrected. Furthermore, we have found at least two occurrences of
incorrect attitude reconstruction for UVOT images of GRBs out of the many
thousands of UVOT images of GRBs processed to date. In such cases the XRT team
will supply an XRT refined position based on telemetered XRT PC-mode data, and
a UVOT-enhanced XRT position from a manual application of the automated
pipeline (using the refined XRT position as the starting position) as soon as
the data allow.

\begin{figure}
\resizebox{\hsize}{!}{\includegraphics[width=8.0cm]{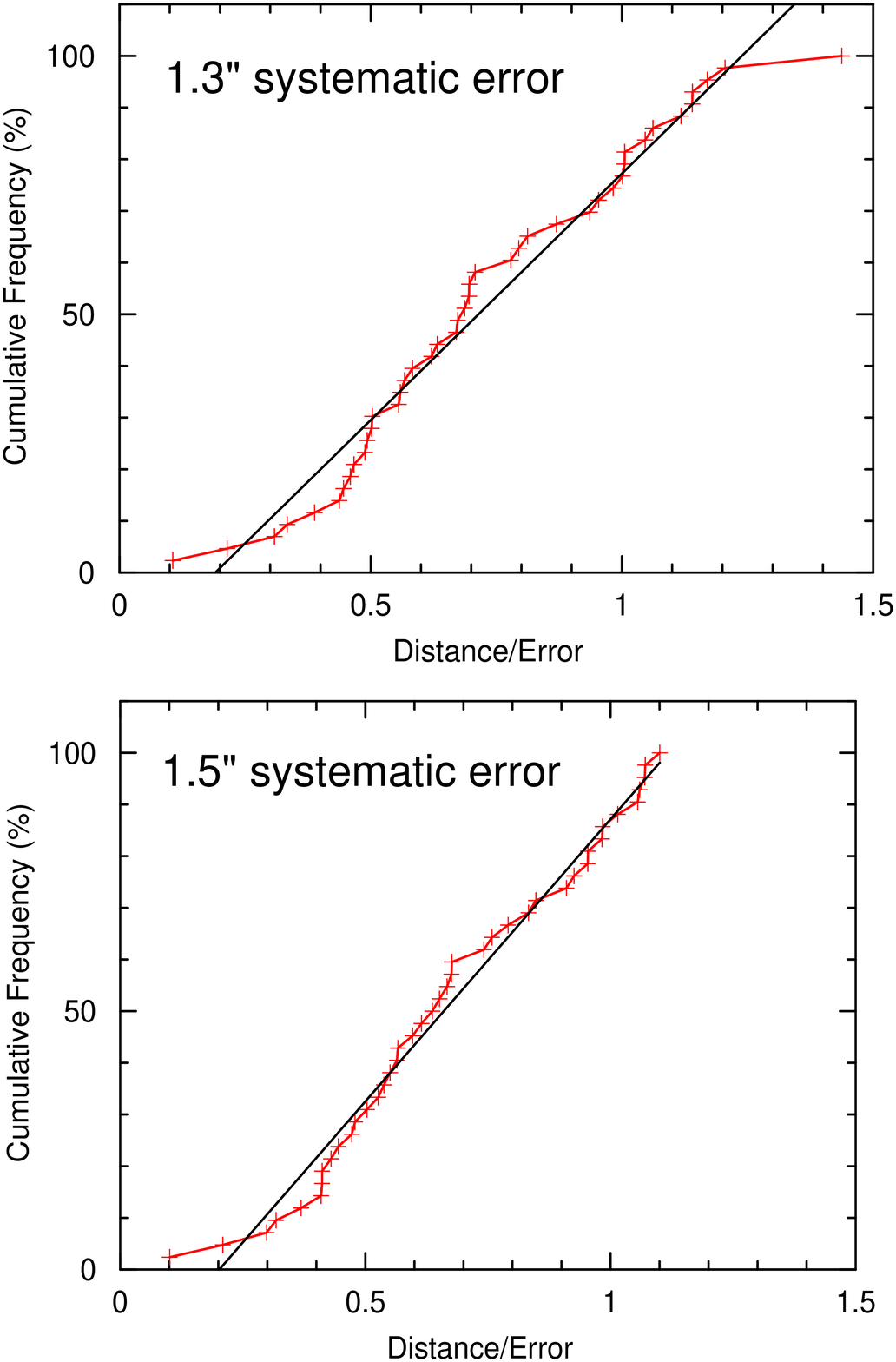}}

\caption{Results of comparing the UVOT-enhanced XRT GRB positions with the
UVOT GRB positions. The upper panel shows that when the 1.3" systematic error
derived from the UVOT-XRT mapping is included in the 90\% error radius, the
positions agree\til80\% of the time.  After increasing the 1.3" systematic to
1.5'' (lower panel), the positions agree 90\% of the time.}

\label{errcheck}
\end{figure}

\section{Validation of results}
\label{test}

In order to validate our procedure, we ran our automated position enhancement
software pipeline for every GRB observed by Swift, yielding 154 enhanced
positions up to GRB~070724A, a success rate of $\approx60$\%. 40 of these GRBS
were also detected by UVOT, for these we calculated the distance between the
UVOT detected position and the UVOT-enhanced XRT position.  Fig.~\ref{precise}
shows the distribution of the offsets between our UVOT-enhanced XRT positions
and UVOT detected positions obtained from the publicly available GCN
notices  for
these 40 GRBs. 90\% of the enhanced XRT positions fall within 2.5 arcseconds
of the UVOT detected position.  We evaluated the uncertainty in the
UVOT-enhanced XRT positions by comparing these positions' offsets from the
UVOT positions divided by the errors of the UVOT-enhanced XRT and UVOT
position errors added in quadrature.

Including the 1.3" uncertainty in the XRT-UVOT mapping as the only systematic
uncertainty we find that the position offset for 90\% of the bursts
is 1.25 times the combined error, suggesting that the systematic error is too
small (Fig.~\ref{errcheck}, upper panel).  Increasing the systematic
error\footnote{Our own analysis of UVOT positions and uncertainties for GRB
optical transients does not result in any additional systematic error
component to that already found.} to 1.5" ensures that the positions agree
within the 90\% errors, 90\% of the time (Fig.~\ref{errcheck}, lower
panel). At present the cause of the extra uncertainty is unknown, and the
increase of our systematic error is a conservative move; it assumes that the
quoted UVOT positions and uncertainties are accurate.
%However, we note that the UVOT positions quoted in the GCN notices are not
%always uniformly generated and consequently their uncertainties are not
%well-quantified; some state explicitly that they have been astrometrically
%corrected and give the systematic error for the correction, others give no
%systematic error, and a small fraction appear to have no astrometric
%correction applied. Some of these pertain to early revolutions of the
%spacecraft (2005), while yet others result from aspect corrections which were
%deemed too small to apply.
It is possible that improved estimates for the UVOT detected source
position and error may reduce the magnitude of this additional systematic
uncertainty\footnote{UVOT position uncertainties are in the range 0.5--1.0'',
thus the bulk of the additional systematic uncertainty in the UVOT-enhanced
XRT positions must originate elsewhere.}. Additionally, as the number of OTs
increases, improvements can be made to the UVOT V-band -- XRT detector mapping
(obviously, additional non-GRB pointings with simultaneous XRT PC mode and
UVOT imaging data can also be used) .

Finally, while concerted effort has been made in determining an
accurate map for the V-band filter, it is our intention to extend this process
to other filters, particularly the White light filter which has the highest
sensitivity and is used early in the filter sequence when the bursts are
brighter.

The UVOT-enhanced XRT positions for 154 GRBs are shown in Table~3. 50\% of the
bursts have position errors below 2.1''; with the X-ray fainter bursts having
larger errors, so that 90\% of all bursts have position errors below 4.2''. The
enhanced positions are automatically relayed to the Gamma Ray Burst community
via GCN circulars, as well as being made available for public scrutiny
on our web-pages located at http://www.swift.ac.uk/xrt\_positions. An example
UVOT V-band image with UVOT-enhanced XRT source positions overlaid is shown in
Fig~\ref{astrom_ximage_all_narrow}.

\section{Summary}

We have constructed an automated pipeline for producing {\it Swift\/} XRT
positions enhanced by UVOT astrometry, which are made available within a few
hours of a GRB trigger. Our position determinations have not only increased
precision ($\approx$ a factor of 2 reduction in error radius) relative to the
refined XRT positions previously in use, but additionally have an accuracy
relative to optically localised positions which is a factor of 2 better. The
positional accuracy obtained using this technique is superior to
astrometrically corrected X-ray positions (obtained by matching X-ray sources
to optical counterparts) at early times (for up to 24 hours), and is comparable
to the accuracy achieved using this technique at later times.

\newpage

\clearpage

\begin{table}
\begin{center}
\caption{UVOT-enhanced XRT positions.} 
\begin{tabular}{lrrr}
\hline\hline
  GRB Name     & RA (J2000)    & Dec (J2000)  & Error$^{\dagger}$ (") \\
               &               &              &            \\
  GRB 070724A  &  01 51 13.98  & $-$18 35 39.8  & 2.2  \\
  GRB 070721B  &  02 12 32.84  & $-$02 11 39.7  & 1.8  \\
  GRB 070721A  &  00 12 39.21  & $-$28 33 00.7  & 2.2  \\
  GRB 070714B  &  03 51 22.29  & $+$28 17 52.2  & 1.7  \\
  GRB 070714A  &  02 51 43.37  & $+$30 14 36.2  & 1.9  \\
  GRB 070704   &  23 38 47.82  & $+$66 15 11.8  & 2.7  \\
  GRB 070628   &  07 41 06.02  & $-$20 16 45.6  & 1.6  \\
  GRB 070621   &  21 35 10.14  & $-$24 49 03.3  & 1.7  \\
  GRB 070616   &  02 08 36.60  & $+$56 56 43.8  & 2.3  \\
  GRB 070615   &  02 57 55.21  & $-$04 26 56.7  & 7.3  \\
  GRB 070612B  &  17 26 54.48  & $-$08 45 03.2  & 3.6  \\
  GRB 070610   &  19 55 09.61  & $+$26 14 06.7  & 4.1  \\
  GRB 070531   &  00 26 58.48  & $+$74 18 46.7  & 1.9  \\
  GRB 070521   &  16 10 38.57  & $+$30 15 21.7  & 1.8  \\
  GRB 070520B  &  08 07 31.07  & $+$57 36 28.9  & 2.5  \\
  GRB 070518   &  16 56 47.83  & $+$55 17 43.0  & 2.2  \\
  GRB 070509   &  15 51 50.51  & $-$78 39 07.1  & 2.5  \\
  GRB 070508   &  20 51 12.02  & $-$78 23 05.0  & 2.0  \\
  GRB 070506   &  23 08 52.34  & $+$10 43 20.3  & 2.9  \\
  GRB 070429A  &  19 50 48.91  & $-$32 24 17.5  & 2.5  \\
  GRB 070420   &  08 04 55.13  & $-$45 33 21.7  & 1.6  \\
  GRB 070419B  &  21 02 49.79  & $-$31 15 48.5  & 1.6  \\
  GRB 070419A  &  12 10 58.83  & $+$39 55 31.5  & 2.3  \\
  GRB 070412   &  12 06 10.11  & $+$40 08 37.1  & 4.6  \\
  GRB 070411   &  07 09 19.78  & $+$01 03 50.9  & 2.6  \\
  GRB 070328   &  04 20 27.60  & $-$34 04 00.6  & 1.8  \\
  GRB 070318   &  03 13 56.69  & $-$42 56 47.3  & 1.7  \\
  GRB 070311   &  05 50 08.21  & $+$03 22 29.6  & 2.7  \\
  GRB 070306   &  09 52 23.25  & $+$10 28 55.5  & 1.6  \\
  GRB 070227   &  08 01 17.00  & $-$46 23 27.8  & 2.0  \\
  GRB 070224   &  11 56 06.65  & $-$13 19 49.6  & 2.0  \\
  GRB 070220   &  02 19 06.86  & $+$68 48 16.1  & 1.7  \\
  GRB 070219   &  17 20 45.91  & $+$69 22 15.6  & 3.2  \\
  GRB 070208   &  13 11 32.76  & $+$61 57 55.2  & 2.4  \\
  GRB 070129   &  02 28 00.89  & $+$11 41 03.3  & 1.8  \\
  GRB 070110   &  00 03 39.30  & $-$52 58 28.6  & 1.7  \\
  GRB 070107   &  10 37 36.49  & $-$53 12 47.5  & 1.6  \\
  GRB 070103   &  23 30 13.74  & $+$26 52 33.2  & 1.8  \\
  GRB 061222A  &  23 53 03.47  & $+$46 31 57.6  & 1.6  \\
  GRB 061202   &  07 02 06.03  & $-$74 41 54.4  & 1.7  \\
  GRB 061201   &  22 08 32.23  & $-$74 34 49.1  & 1.7  \\
  GRB 061126   &  05 46 24.67  & $+$64 12 40.0  & 1.6  \\
  GRB 061122   &  20 15 19.79  & $+$15 31 02.3  & 2.0  \\
  GRB 061110A  &  22 25 09.81  & $-$02 15 27.9  & 1.9  \\
  GRB 061021   &  09 40 36.17  & $-$21 57 05.3  & 1.6  \\
  GRB 061019   &  06 06 30.92  & $+$29 34 12.8  & 2.3  \\
  GRB 061007   &  03 05 19.70  & $-$50 30 02.7  & 1.9  \\
  GRB 061006   &  07 24 07.88  & $-$79 11 53.2  & 3.2  \\
  GRB 061004   &  06 31 10.80  & $-$45 54 24.3  & 1.7  \\
  GRB 060929   &  17 32 28.93  & $+$29 50 06.7  & 2.0  \\
  GRB 060927   &  21 58 12.01  & $+$05 21 49.8  & 2.1  \\
  GRB 060926   &  17 35 43.75  & $+$13 02 14.3  & 4.5  \\
  GRB 060923B  &  15 52 46.64  & $-$30 54 14.7  & 2.5  \\
  GRB 060923A  &  16 58 28.12  & $+$12 21 38.8  & 2.1  \\
  GRB 060919   &  18 27 41.76  & $-$51 00 52.4  & 2.3  \\
  GRB 060912A  &  00 21 08.09  & $+$20 58 19.1  & 1.7  \\
  GRB 060908   &  02 07 18.43  & $+$00 20 32.8  & 1.6  \\
\hline

\end{tabular}
\end{center}
\end{table}

\addtocounter{table}{-1}

\begin{table}
\begin{center}
\caption{Continued - UVOT enhanced XRT positions.}
\begin{tabular}{lrrr}
\hline\hline
  GRB Name     & RA (J2000)    & Dec (J2000)  & Error (") \\
       &     &       &            \\
  GRB 060906   &  02 43 00.83  & $+$30 21 42.6  & 2.1  \\
  GRB 060904B  &  03 52 50.57  & $-$00 43 29.9  & 2.5  \\
  GRB 060825   &  01 12 21.54  & $+$55 46 41.4  & 7.3  \\
  GRB 060814   &  14 45 21.34  & $+$20 35 09.6  & 1.6  \\
  GRB 060813   &  07 27 35.33  & $-$29 50 49.4  & 1.8  \\
  GRB 060807   &  16 50 02.62  & $+$31 35 30.2  & 1.6  \\
  GRB 060805A  &  14 43 43.35  & $+$12 35 12.3  & 2.8  \\
  GRB 060804   &  07 28 49.27  & $-$27 12 56.1  & 1.7  \\
  GRB 060801   &  14 12 01.32  & $+$16 58 54.4  & 1.9  \\
  GRB 060729   &  06 21 31.81  & $-$62 22 12.1  & 1.7  \\
  GRB 060719   &  01 13 43.72  & $-$48 22 50.9  & 1.9  \\
  GRB 060714   &  15 11 26.45  & $-$06 33 59.3  & 1.7  \\
  GRB 060708   &  00 31 13.72  & $-$33 45 34.1  & 1.9  \\
  GRB 060707   &  23 48 19.11  & $-$17 54 17.6  & 1.7  \\
  GRB 060614   &  21 23 32.06  & $-$53 01 36.2  & 1.9  \\
  GRB 060607A  &  21 58 50.46  & $-$22 29 47.3  & 1.6  \\
  GRB 060605   &  21 28 37.32  & $-$06 03 30.7  & 1.7  \\
  GRB 060604   &  22 28 55.36  & $-$10 54 55.4  & 5.2  \\
  GRB 060526   &  15 31 18.31  & $+$00 17 06.2  & 1.8  \\
  GRB 060522   &  21 31 44.86  & $+$02 53 10.1  & 1.8  \\
  GRB 060512   &  13 03 05.70  & $+$41 11 26.5  & 1.9  \\
  GRB 060510B  &  15 56 29.37  & $+$78 34 11.6  & 2.0  \\
  GRB 060507   &  05 59 50.41  & $+$75 14 56.0  & 2.6  \\
  GRB 060502B  &  18 35 44.79  & $+$52 37 55.7  & 6.3  \\
  GRB 060502A  &  16 03 42.58  & $+$66 36 01.5  & 1.8  \\
  GRB 060428B  &  15 41 25.77  & $+$62 01 29.8  & 2.4  \\
  GRB 060428A  &  08 14 10.77  & $-$37 10 11.7  & 1.7  \\
  GRB 060427   &  08 17 04.29  & $+$62 40 17.0  & 2.3  \\
  GRB 060418   &  15 45 42.70  & $-$03 38 21.0  & 2.0  \\
  GRB 060413   &  19 25 07.82  & $+$13 45 30.1  & 1.6  \\
  GRB 060319   &  11 45 32.91  & $+$60 00 39.2  & 2.0  \\
  GRB 060313   &  04 26 28.41  & $-$10 50 41.2  & 2.0  \\
  GRB 060306   &  02 44 22.74  & $-$02 08 55.1  & 3.5  \\
  GRB 060223A  &  03 40 49.50  & $-$17 07 48.7  & 4.2  \\
  GRB 060218   &  03 21 39.65  & $+$16 52 01.0  & 2.1  \\
  GRB 060211A  &  03 53 32.70  & $+$21 29 19.2  & 4.1  \\
  GRB 060210   &  03 50 57.31  & $+$27 01 33.8  & 1.6  \\
  GRB 060130   &  15 16 30.53  & $-$36 50 07.4  & 2.1  \\
  GRB 060206   &  13 31 43.51  & $+$35 03 02.0  & 2.7  \\
  GRB 060204B  &  14 07 15.05  & $+$27 40 36.9  & 2.0  \\
  GRB 060203   &  06 54 04.00  & $+$71 48 39.3  & 3.5  \\
  GRB 060202   &  02 23 22.92  & $+$38 23 03.6  & 1.8  \\
  GRB 060123   &  11 58 47.75  & $+$45 30 51.4  & 4.4  \\
  GRB 060121   &  09 09 52.14  & $+$45 39 47.7  & 2.4  \\
  GRB 060116   &  05 38 46.24  & $-$05 26 13.9  & 3.0  \\
  GRB 060115   &  03 36 08.27  & $+$17 20 41.8  & 2.5  \\
  GRB 060111B  &  19 05 42.55  & $+$70 22 32.2  & 2.2  \\
  GRB 060111A  &  18 24 49.22  & $+$37 36 15.7  & 4.6  \\
  GRB 060109   &  18 50 43.65  & $+$31 59 27.2  & 2.3  \\
  GRB 060105   &  19 50 00.74  & $+$46 20 56.6  & 1.7  \\
  GRB 051221A  &  21 54 48.24  & $+$16 53 22.7  & 3.6  \\
  GRB 051211B  &  23 02 41.13  & $+$55 04 48.8  & 2.5  \\
  GRB 051109B  &  23 01 50.29  & $+$38 40 47.2  & 2.3  \\
  GRB 051109A  &  22 01 15.30  & $+$40 49 22.7  & 1.7  \\
  GRB 051028   &  01 48 14.99  & $+$47 45 10.0  & 4.2  \\
  GRB 051021A  &  01 56 36.25  & $+$09 04 01.0  & 2.2  \\
  GRB 051016B  &  08 48 27.97  & $+$13 39 19.7  & 2.4  \\
\hline

\end{tabular}
\end{center}
\end{table}
\addtocounter{table}{-1}

\begin{table}
\begin{center}
\caption{Continued - UVOT enhanced XRT positions.}
\begin{tabular}{lrrr}
\hline\hline
  GRB Name     & RA (J2000)    & Dec (J2000)  & Error (") \\
       &     &       &            \\
  GRB 051016A  &  08 11 16.93  & $-$18 17 55.4  & 4.3  \\
  GRB 051012   &  18 02 28.59  & $-$52 36 49.0  & 1.8  \\
  GRB 051008   &  13 31 29.45  & $+$42 05 52.7  & 2.0  \\
  GRB 050922C  &  21 09 32.97  & $-$08 45 30.6  & 1.7  \\
  GRB 050922A  &  18 04 21.99  & $-$32 10 51.1  & 4.6  \\
  GRB 050922B  &  00 23 13.68  & $-$05 36 17.0  & 2.6  \\
  GRB 050918   &  17 51 12.14  & $-$25 34 39.0  & 3.3  \\
  GRB 050915A  &  05 26 45.14  & $-$28 01 01.1  & 2.8  \\
  GRB 050908   &  01 21 50.92  & $-$12 57 17.8  & 4.6  \\
  GRB 050904   &  00 54 51.13  & $+$14 05 06.0  & 1.9  \\
  GRB 050827   &  04 17 09.59  & $+$18 12 02.0  & 2.6  \\
  GRB 050826   &  05 51 01.65  & $-$02 38 34.9  & 4.7  \\
  GRB 050822   &  03 24 27.37  & $-$46 02 00.8  & 2.2  \\
  GRB 050820A  &  22 29 38.12  & $+$19 33 37.1  & 1.8  \\
  GRB 050819   &  23 55 01.53  & $+$24 51 41.2  & 3.5  \\
  GRB 050815   &  19 34 22.98  & $+$09 08 46.6  & 5.1  \\
  GRB 050814   &  17 36 45.30  & $+$46 20 21.6  & 2.1  \\
  GRB 050813   &  16 07 56.77  & $+$11 14 56.3  & 1.7  \\
  GRB 050803   &  23 22 37.91  & $+$05 47 09.4  & 1.7  \\
  GRB 050802   &  14 37 05.80  & $+$27 47 11.5  & 2.3  \\
  GRB 050730   &  14 08 17.21  & $-$03 46 19.3  & 1.6  \\
  GRB 050724   &  16 24 44.07  & $-$27 32 26.5  & 2.9  \\
  GRB 050717   &  14 17 24.49  & $-$50 32 00.8  & 2.3  \\
  GRB 050716   &  22 34 20.40  & $+$38 41 07.0  & 3.7  \\
  GRB 050714B  &  11 18 47.66  & $-$15 32 48.4  & 3.6  \\
  GRB 050713A  &  21 22 09.34  & $+$77 04 29.6  & 1.8  \\
  GRB 050712   &  05 10 48.29  & $+$64 54 48.6  & 2.4  \\
  GRB 050603   &  02 39 56.96  & $-$25 10 54.2  & 1.7  \\
  GRB 050509C  &  12 52 21.03  & $-$44 26 00.9  & 2.5  \\
  GRB 050505   &  09 27 03.30  & $+$30 16 23.8  & 1.6  \\
  GRB 050504   &  13 24 01.29  & $+$40 42 14.7  & 3.6  \\
  GRB 050502B  &  09 30 10.13  & $+$16 59 47.5  & 1.9  \\
  GRB 050416A  &  12 33 54.55  & $+$21 03 27.0  & 1.9  \\
  GRB 050408   &  12 02 17.31  & $+$10 51 09.5  & 2.1  \\
  GRB 050319   &  10 16 47.93  & $+$43 32 55.3  & 1.8  \\
  GRB 050318   &  03 18 50.95  & $-$46 23 44.7  & 2.2  \\
  GRB 050315   &  20 25 54.07  & $-$42 36 01.5  & 2.0  \\
  GRB 050306   &  18 49 14.58  & $-$09 09 06.9  & 5.3  \\
  GRB 050219B  &  05 25 16.02  & $-$57 45 27.8  & 2.8  \\
  GRB 050124   &  12 51 30.70  & $+$13 02 41.3  & 4.7  \\
\hline

\end{tabular}
\end{center}
$\dagger$All errors 90\% confidence radius.
\end{table}

\clearpage

\clearpage

%\end{appendix}

\end{document}